\documentclass[showpacs,showkeys,amsmath,amssymb,fleqn,preprint]{revtex4}
\usepackage{graphicx,amsmath,amsfonts,amssymb,float,bm,epsfig,color}

\newcommand{\newc}{\newcommand}

\newc{\be}{\begin{equation}}
\newc{\ee}{\end{equation}}
\newc{\bea}{\begin{eqnarray}}
\newc{\eea}{\end{eqnarray}}
\newc{\beas}{\begin{eqnarray*}}
\newc{\eeas}{\end{eqnarray*}}

\newc{\pardt}{\partial_{t}}
\newc{\pardxi}{\partial_{i}}
\newc{\pardts}{\partial_{t^{*}}}
\newc{\pardxis}{\partial_{i^{*}}}
\newc{\pardxj}{\partial_{j}}
\newc{\pardxk}{\partial_{k}}
\newc{\pard}{\partial}
\newc{\ti}{\tilde}

\newc{\s }{\overline}

\newc{\sect}{\section}
\newc{\subs}{\subsection}

\newc{\defi}{\definition}
\newc{\prop}{\proposition}
\newc{\rem}{\remark}
\newc{\lem}{\lemma}
\newc{\exa}{\example}
\newc{\theo}{\theorem}
\newc{\coro}{\corollary}
\newc{\post}{\postulate}
\newc{\state}{\statement}

\def\d{{\textrm{d}}}

\begin{document}
\baselineskip0.5cm
\renewcommand {\theequation}{\thesection.\arabic{equation}}

\title{Transition to ballistic regime for  heat transport in helium II}

\author{M.~Sciacca\footnote{Corresponding author.}}
    \email{michele.sciacca@unipa.it}
    \affiliation{Dipartimento Scienze Agrarie e Forestali, Universit\`a di Palermo, Viale delle Scienze, 90128 Palermo, Italy}
    \affiliation{Departament de F\'isica, Universitat Aut\`onoma de Barcelona, 08193 Bellaterra, Catalonia, Spain}
\author{A. Sellitto}
    \email{ant.sellitto@gmail.com}
    \affiliation{Dipartimento di Matematica, Informatica ed Economia,  Universit\`a della Basilicata, Campus Macchia Romana, 85100, Potenza, Italy}
\author{D. Jou}
    \email{david.jou@uab.cat}
    \affiliation{Departament de F\'isica, Universitat Aut\`onoma de Barcelona, 08193 Bellaterra, Catalonia, Spain}
    \affiliation{Institut d'Estudis Catalans, Carme 47, Barcelona 08001, Catalonia, Spain}



\begin{abstract}
The size-dependent and  flux-dependent effective thermal conductivity of narrow capillaries filled with superfluid helium is analyzed from a thermodynamic continuum perspective. The classical Landau evaluation of the effective thermal conductivity of quiescent superfluid, or the Gorter-Mellinck regime of turbulent superfluids, are extended to describe the transition to ballistic regime in narrow channels wherein the radius $R$ is comparable to (or smaller than) the phonon mean-free path $\ell$ in superfluid helium. To do so we start from an extended equation for the heat flux incorporating non-local terms, and take into consideration a heat slip flow along the walls of the tube. This leads from an effective thermal conductivity proportional to $R^2$ (Landau regime) to another one proportional to $R\ell$ (ballistic regime).
We consider two kinds of flows: along cylindrical pipes and along two infinite parallel plates.
\end{abstract}


\maketitle

\section{Introduction}
Understanding the size dependence of the effective thermal conductivity of systems with size comparable to the mean free path of heat carriers is a central topic in transport theory in nanosystems\;\citep{Tzou-book(1997), Zhang-book(2007), Chen-book(2005), JCC-S, Volz_book(2010)}.
In such systems, the effective thermal conductivity is much reduced with respect to the bulk thermal conductivity. In recent years this topic has been much studied in solid systems either from microscopic perspective based on kinetic theory or molecular dynamics\;\citep{Struchtrup_book(2005), Sharipov_PRE69(2004), Cahill_JAP93(2003), Hochbaum_Nat451(2008)}, or from mesoscopic generalizations of the transport equations\;\cite{Lebon_JNT39(2014), Cimmelli_JNT34(2009), Vazquez_JAP104(2009), WangPLA376(2012), WangPLA374(2010), Lebon_PLA376(2012), Cimmelli_PRB82(2010), Sellitto_JAP107(2010), AlvJouSel1, Tzou_JTS49(2010)}. 
The aim of this paper is to deal with this problem in superfluid helium in narrow channels. For low temperature helium, below some 0.7 K, the phonon mean free path is of the order of 0.5 mm ($5 \times 10^5$ nm), or even longer. Thus, superfluid helium is an interesting system for these analyses, becuase it does not require truly nanometric channels.

Heat tranport in superfluid helium is much richer and diverse than in semiconductors. In this paper we use a simple thermodynamical model to describe the transtion from diffusive to ballistic heat transport in helium II. We compare it with an analogous description in solid nanowires. 

The layout of the paper is the following. In Section\;\ref{2} we review the three main regimes for heat transport in superfluid helium (the Landau regime, the Gorter-Mellinck regime and the ballistic regime).   In Section~\ref{section2} we briefly review the derivation of the effective thermal conductivity  in  the Landau regime in terms of the one-fluid model. In Section~\ref{section3} we describe the transition from laminar diffusive regime to balistic regime. Section~\ref{section4} is devoted to the corresponding transition in the presence of turbulence. Final comments are given in Section~\ref{Concl}.

\section{Heat transport features of superfluid helium}\label{2}
\setcounter{equation}{0}
Heat transport in superfluid helium (He II) in narrow channels has been a  topic of interest since 1950's, due to the peculiar ability of superfluid helium to flow along very narrow channels~\cite{ArpJAP40(1969),Mendelsohn,VASbook,Bertman_Cry8(1968),Benin_PRB18(1978),Greywall_PRB23(1981),Maris_PRA7(1973)}. Currently, this topic is again of interest for the cooling of nanosystems. Heat transport in superfluid helium has several special features related to the relative presence of phonons and rotons, the laminar or turbulent flow, and the relation between phonon mean-free path (mfp) and the radius of the container, which gives to this topic a rich phenomenology, much wider than that of heat flow in solids.

We summarize these features in a sketch of the essential observations:
    \begin{enumerate}
        \item \textbf{Landau regime:} When the phonon mfp is short as compared with the radius of the pipe, and the heat-flux value is low enough, there is viscous laminar flow of the normal component of helium (carrying the heat flow) which is described by the following expression~\cite{CritchlowJAP40(1969),Mendelsohn,LL,VASbook}:
                \be\label{Delta_Landau}
                    \nabla T=-\frac{8 \eta}{\pi R^4 S^2 T} \dot Q
                \ee
            where $R$ is radius of the pipe, $\eta$ is the viscosity of the normal component, $S$ the entropy per unit volume, $T$ absolute temperature, $\dot Q$ the total heat current across the pipe (namely, $\dot Q=\pi R^2 q$, $q$ being the local heat flux, or the heat flowing per unit time and unit area), and $\nabla T$ the local temperature gradient  along the tube. The coefficient relating $\nabla T$ and $\dot Q$ in Eq.~\eqref{Delta_Landau} is the thermal resistance of the pipe.
        \item\textbf{Gorter-Mellinck regime:} The laminar flow, corresponding to Landau regime, breaks down for sufficiently high values of the heat flux. In this case, quantized vortices appear and contribute to the thermal resistance~\cite{Donnelly,Barenghi_libro,TsuKobTak,NemPR2013}, because of the frictional force between the normal component and the quantized vortices. The relation between $\nabla T$ and $\dot Q$ for fully-developed turbulence is given by~\cite{ArpJAP40(1969),Mendelsohn}
                \be\label{Delta_Gorter}
                    \nabla T=-\left(\frac{C \rho_n }{S^4 T^3 \rho_s^3}\right)\left(\frac{\dot Q}{\pi R^2}\right)^3
                \ee
            with $\rho_n$ and $\rho_s$ as the mass densities of normal component and superfluid component, and $C$ a numerical constant. In this regime, the friction between the normal component (carrying the heat flow) and the quantized vortex tangle provides the main part of thermal resistance, in contrast to the purely viscous resistance in Landau regime.
        \item\textbf{Ballistic regime:} The phonon mfp $\ell$ increases when temperature is lowered (for low temperatures, for example, it behaves as $T^{-4.3}$~\cite{Bertman_Cry8(1968)}), in such a way that for sufficiently low temperatures, it becomes comparable to (or higher than) the radius of the pipe (this happens below some $0.7\operatorname{K}$ for $R$ of the order of $0.5\operatorname{mm}$, but it would occur  for higher temperature if smaller diameters are considered)~\cite{Bertman_Cry8(1968), Greywall_PRB23(1981)}. In this case, the predominant collisions are not the phonon-phonon collisions, but the phonon-walls collisions~\cite{Bertman_Cry8(1968)}. Thus, in these situations the walls play a crucial role. The expression relating $\nabla T$ and $\dot Q$ in this case is~\cite{Bertman_Cry8(1968),Greywall_PRB23(1981)}
                \be\label{Delta_f}
                    \nabla T=-\left(\frac{3}{2C_vv\pi R^3}\right)\left(\frac{f}{2-f}\right)\dot Q
                \ee
            with $C_v$ being the phonon specific heat, $v$ being the modulus of the phonon speed, and $f$ is the fraction of phonons undergoing diffuse scattering from the tube walls (in contrast to those undergoing specular scattering).
    \end{enumerate}

From a practical point of view, one of the essential  differences between these regimes is the  dependence of $\dot Q l/\Delta T$ ($l$ being the length of the duct) with the radius, which is proportional to $R^4$ in the Landau regime, and to $R^3$ in the ballistic regime, and with the heat flux, which is proportional to $\dot Q^2$ in the turbulent regime and independent of $\dot Q$ in the laminar and the ballistic regimes.

From a non-equilibrium thermodynamics perspective, it is convenient to have a general wide enough model able to describe these three different regimes, and the transition between them. Such transitions are characterized by the ratios $\displaystyle\frac{\ell}{d}$, $\displaystyle\frac{L^{-1/2}}{d}$ and $\displaystyle\frac{L^{-1/2}}{\ell}$, with $L$ as the vortex length density, and $d=2R$ as the diameter of the tube, or the distance between the plates in the rectangular channel. The transition between the Landau regime ($\ell\ll d \ll L^{-1/2}$) and the turbulent regime ($\ell\ll L^{-1/2} \ll d$) has been recently considered in Ref.~\cite{Sciacca_2013} using  an equation for the heat flux $q$ and another one for the vortex length density $L$~\cite{monjouNSP2007}, which is a function of $q$, and plays a central role in the transition from Eq.~\eqref{Delta_Landau} to Eq.~\eqref{Delta_Gorter}. This is an important transition because it leads to a big increase of thermal resistance, leading to a high loss of efficiency in cooling, and may eventually lead to a burnout of the system if  the fluid helium crosses the lambda temperature (some $2.2\operatorname{K}$) and it becomes a normal viscous fluid, instead of a superfluid.

The aim of the present paper is to generalize such model by allowing it to describe the transition to the ballistic regime. More precisely, our aim is to consider  the transition from the Landau regime to the ballistic regime, which occurs for $d\ll \ell \ll L^{-1/2}$ or $d\ll L^{-1/2} \ll \ell$ (absence of vortices), and the transition from the turbulent Gorter-Mellinck regime to the ballistic regime, which means  $L^{-1/2}\ll d \ll \ell$ or $L^{-1/2}\ll \ell \ll d$.

We will describe these transitions in terms of the one-fluid model of He II with the heat flux as an internal variable~\cite{Mongiovi1993}, which was used in Ref.~\cite{Sciacca_2013}, by complementing it with  a constitutive equation for a slip heat flux along the walls. Indeed, in the collisional  situation the standard  two-fluid model is less satisfactory than in usual situations, because the viscous model is related to diffusive behavior but not to the ballistic one.

\section{Effective thermal conductivity of superfluid helium}\label{section2}
\setcounter{equation}{0}
In this section we briefly review the macroscopic derivation of the effective thermal conductivity of He II in the laminar regime, i.e., in the absence of vortices, along a cylindrical duct or between two inifinte plates (as simple example to emulate a rectangular channel with high aspect ratio). We describe heat transport in terms of the one-fluid model of Extended Thermodynamics~\cite{Mongiovi1993}, and compare with Tisza-Landau two-fluid model\;\cite{Bertman_Cry8(1968),Benin_PRB18(1978),Greywall_PRB23(1981),Maris_PRA7(1973)}. If liquid helium is globally at  rest, the motion of the normal component is compensated by an opposite flow of the superfluid component, in such a way that the net velocity of the total system vanishes, i.e., there is no net mass flow. This requires that at any time, one has $\rho_s\bar v_s+\rho_n \bar v_n =0$, where $\bar v_s$ and $\bar v_n$ are the average velocities of the superfluid and the normal components on the transversal section of the tube. This situation is called {\it counterflow} in literature on liquid helium II~\cite{Donnelly, Barenghi_libro,NemPR2013,TsuKobTak}, and the relevant quantity here is the so-called counterflow velocity $v_{ns}$, defined as
    \[
        v_{ns}=\bar v_{n}-\bar  v_{s}=\frac{\rho}{\rho_s}{\bar v}_{n}.
    \]

The second equality of the former equation directly follows from the mentioned condition of vanishing mass flow, namely, $\rho_n \bar v_{n}+\rho_s \bar v_{s}=0$. Note for further use that the heat flow is given by
    \[
        \bar q= ST \bar{v}_n=\frac{\rho_s}{\rho}ST{v}_{ns}.
    \]

In the situation considered here, there is no net flow of matter along the channel, i.e., we consider this cylinder as a pore connecting a body at a temperature $T$ with a refrigerating helium heat bath with no heat exchange across the lateral walls of the capillary.

According to the one-fluid model with the local heat flux $q$ as internal variable, the dynamical equations in the stationary situation for zero net mass flow are
	\begin{equation}\label{eq:heat_flux}
		\left\{
			\begin{array}{lll}
                \vspace{0.2cm}\displaystyle
                \frac{\pard q_j}{\pard x_j}=0,\\
                \vspace{0.2cm}\displaystyle
				\frac{\pard }{\pard x_j}\left(p\delta_{ij}+m_{<ij>}\right)=0,\\
				\vspace{0.2cm}\displaystyle
			    m_{<ij>}= 2\beta T\lambda_2\frac{\pard q_{<i}}{\pard x_{j>}},\\
			    \vspace{0.2cm}\displaystyle
  	            \lambda_1 \frac{\pard T}{\pard x_i}-\beta T^2\lambda_1\frac{\pard }{\pard x_j}m_{<ij>}=\sigma^q_i,
			\end{array}
			\right.
	\end{equation}
where the coefficient $\beta$ may be set $\beta=-\left(ST^2\right)^{-1}$~\cite{Mongiovi1993}, $\sigma^q$ is the production term of the heat flux, $p$ is pressure, $m_{<ij>}$ the flux of heat flux, and $\lambda_1$ and $\lambda_2$ can be interpreted as the heat conductivity and the shear viscosity, respectively, when applied to a classical fluid with $\sigma^q=- {\bf q}$~\cite{Mongiovi1993}. In particular the phenomenological coefficient $\lambda_2$ may be set equal to the shear viscosity $\eta$ of the normal component, i.e., $\lambda_2=\eta$ in Eq.~\eqref{eq:heat_flux}.

In this section we assume that $\sigma^q=- {\bf q}$ while a more general assumption is required to take into account of the presence of vortices~\cite{JCC-S,Sciacca_2013}. In these equations, the time derivatives of the corresponding quantities have been neglected because we are interested in steady state situations.

After some trivial substitutions, and neglecting the nonlinear terms, as for instance terms like $\displaystyle 2\lambda_2 \frac{\pard q_{<i}}{\pard x_{j>}}\frac{\pard }{\pard x_j}\left(\beta T\right)$, Eqs.~\eqref{eq:heat_flux}b and~\eqref{eq:heat_flux}d, respectively, become
    \begin{subequations}
        \label{eq:heat_flux2}
        \begin{align}
            \vspace{0.2cm}\displaystyle
            &\nabla p-\frac{\eta}{ST}\nabla^2{\bf q}=0,\label{eq:heat_flux2a}\\
            &S \nabla T-\nabla p=-\frac{S}{\lambda_1}{\bf q}.\label{eq:heat_flux2b}
        \end{align}
    \end{subequations}

The thermal conductivity $\lambda_1$ is related to the velocity of second sound $w_2$ by the relation $\zeta=\lambda_1/\tau_1=w_2^2\rho C_v$, where $\tau_1$ is the relaxation time of the heat flux. Both $\tau_1$ and $\lambda_1$ are very high in superfluid helium, but their ratio is finite.

Let's investigate the case in which the right-hand side of Eq.~\eqref{eq:heat_flux2b} is negligible when compared to the left-hand side, namely, for instance when $\lambda_1$ is high enough, as it is experimentally observed. In this case, Eqs.~\eqref{eq:heat_flux2} become:
    \begin{subequations}\label{eq:heat_flux2bis}
        \begin{align}
            \vspace{0.2cm}\displaystyle
            &\nabla  p-\frac{\eta}{ST} \nabla^2{\bf q}=0,\label{eq:heat_flux2bisa}\\
            &S \nabla T-\nabla p=0.\label{eq:heat_flux2bisb}
        \end{align}
    \end{subequations}

\subsection{Cylindrical channel}
By means of Eq.~\eqref{eq:heat_flux2bisb}, Eq.~\eqref{eq:heat_flux2bisa} becomes
	\be\label{eq:heat_flux3m}
		-\nabla T+\frac{\eta}{S^2 T}\nabla^2{\bf q}=0.
	\ee

Let's consider that temperature $T$ is homogeneous in any transversal section of the pipe and that it changes along the longitudinal axis of the pipe. In this case, in a given transversal section, the heat flux ${\bf q}$ depends only on the radius $r$ of the pipe, and the same procedure adopted in classical hydrodynamics for Poiseuille flow leads to
 	\be\label{eq:heat_flux3}
 		q\left(r\right)=\frac{R^2S^2T\nabla T}{4\eta}\left(\frac{r^2}{R^2}-1\right)+q_w,
	\ee
where $R$ is the radius of the pipe, and the constant of integration $q_w$ is the slip heat flux along the wall (which is zero for viscous fluid, but which may be different from zero in rarefied systems, as it will be commented below). We keep it for the moment, for the sake of completeness and for further use below.

In this situation and when the contribution along the walls $q_w$ may be neglected, the total heat flow across any transversal section of the tube will be
	\be\label{eq:heat_flux4}
		\dot Q=\int_0^{2\pi}\int_0^Rq\left(r\right)r \d r\d \theta=-\left(\frac{\pi R^4S^2T}{8\eta}\right)\frac{\d T}{\d x},
 	\ee
which can be integrated along the pipe from $0$ to $l$ in order to have that the total thermal conductance along the whole tube is
	\be\label{eq:heat_flux5}
		\dot Q l=\int_{T_1}^{T_2}\pi R^2\frac{R^2S^2T}{8\eta}\d T=\left(\frac{\pi R^4S^2}{8\eta}\right)T_M\Delta T,
 	\ee
where $T_M=\left(T_1+T_2\right)/2$ and $\Delta T=\left(T_2-T_1\right)$, with $T_2$ larger than $T_1$. For the sake of clarity, we note that in deriving Eq.~\eqref{eq:heat_flux5} we supposed that the transversal section at $x=0$ is kept at $T_1$, whereas the transversal section at $x=l$ is kept at $T_2$. According to the classical Fourier law, form Eq.~\eqref{eq:heat_flux5} we obtain the following effective thermal conductivity
	\be\label{eq:conductivity1}
		K_{\text{eff}}=\left(\frac{\dot Q}{\pi R^2}\right)\left(\frac{l}{\Delta T}\right)=\frac{R^2T_MS^2}{8\eta},
	\ee
which corresponds to Eq.~\eqref{Delta_Landau}. We explicitly note that the integrand $\displaystyle k=\frac{R^2S^2 T}{8 \eta}$ in Eq.~\eqref{eq:heat_flux5} is also referred as the thermal conductivity in Ref.~\cite{Bertman_Cry8(1968)}. However, if $T_1\approx T_2$, then $K_{\text{eff}}$ in Eq.~\eqref{eq:heat_flux5} reduces to that value of $k$, because  $T_M\approx T_1\approx T_2$.

\subsection{Ractangular channel}
Here, instead, we take into account a channel bounded by two infinite parallel plates separated by a distance  $d$ and filled   by superfluid helium. The heat flows parallel to the plates and we choose the axis $x$ in this direction, whereas the axis $z$ is orthogonal to the plates. Moreover we assume that pressure and temperature depend only on the variable $x$.

By the above hypothesis, we have that ${\bf q}\left(z\right)=\left(q\left(z\right),0,0\right)$ with $-d/2\leq z \leq d/2$, whereas $T=T\left(x\right)$ and $p=p\left(x\right)$. In this case Eq.~\eqref{eq:heat_flux2bisa} becomes
	\be\label{eq:heat_flux12bis}
		\frac{\d}{\d x}  p-\frac{\eta}{ST}\frac{\d^2}{\d z^2} q=0
	\ee
which can be easily integrated getting
	\be\label{eq:soluz}
		q\left(z\right)=\frac{ST}{8\eta} \nabla p\left(4 z^2-d^2\right)+q_w.
	\ee

In the Landau regime the phonon mfp is shorter than the channel size $d$ and the value $q_w$ at the wall has to be zero because of the viscous non-slip conditions. From Eq.~\eqref{eq:soluz} it follows that the heat flow across a generic transversal section of width $a$ and height $d$ is
	\be\label{eq:heat_flux17}
		\dot Q= a \int_{-d/2}^{d/2} \frac{ST}{8\eta} \nabla p\left(4 z^2-d^2\right) \d z =
		-\frac{a STd^3}{12 \eta} \nabla p
	\ee
which, by means of the relation $S\nabla T=\nabla p$, becomes
	\be\label{eq:heat_flux18}
		\dot Q= -\frac{a S^2Td^3}{12 \eta} \nabla T.
	\ee

The heat flow\;\eqref{eq:heat_flux18} can be integrated over a distance $l$ of the channel having temperature $T_2 >T_1$ at the edge. The result is
	\be\label{eq:heat_flux18bis}
		l\dot Q=\frac{a S^2T_Md^3}{12\eta}\Delta T
	\ee
where $\displaystyle T_M=\frac{T_1+T_2}{2}$ and $\Delta T=T_1-T_2$. The effective thermal conductivity arising from Eq.~\eqref{eq:heat_flux18bis} is
\be\label{eq:conductivity1a}
		K_{\text{eff}}= \frac{d^2T_MS^2}{12\eta}.
	\ee

In closing the present section we observe that to describe the transition to the Gorter-Mellinck regime the term $\sigma^q$ in Eq.~\eqref{eq:heat_flux}d has to be generalized to the form $\displaystyle\sigma^q=-\mathbf{q}-\tau_1 K L\mathbf{q}$, with $K$ being a constant proportional to the Hall-Vinen friction coefficient due to the phonon-vortex interactions. In this case, an evolution equation for $L$ has also to be introduced, in order to express it in terms of $\mathbf{q}$ in steady-states situations. This has been done in Ref.~\cite{Sciacca_2013}, and will not be represented here.

\section{Transition from Landau to ballistic regime}\label{section3}
\setcounter{equation}{0}
Many researchers in the past (see, for instance, Refs.~\cite{Bertman_Cry8(1968), Greywall_PRB23(1981)}) have focused own attention on the behavior of heat inside flowing superfluid helium, which depends strongly on the temperature of interest. At temperature below $1\operatorname{K}$ (for the dimension of the channels considered in the experiments) the excitations (rotons and phonons) cannot be considered as a normal fluid, but as a rarefied gas because their mfp is of the same order of the channel's diameter. Moreover, for temperatures lower than $0.7\operatorname{K}$~\cite{Greywall_PRB23(1981)} heat is carried by phonons, and the presence of rotons is negligible.

To describe the transition from the Landau regime to the ballistic regime we consider, in analogy with the analysis in Ref.~\cite{AlvJouSel1} for phonons in solids, the role of the slip heat flow $q_w$ along the walls appearing in Eq.~\eqref{eq:heat_flux3} and~\eqref{eq:soluz}. In the previous section we have set  $q_w=0$, because of the viscous character of the normal component. Though the superfluid component may slip along the walls, the normal component cannot do the same. Since the counterflow condition requires $\rho_n \bar v_n +\rho_s \bar v_s=0,$ a slip heat flow along the wall in the Landau regime is not possible, because it would imply a mass flow (because ${\bf \bar{v}}_n$ should be zero but not ${\bf \bar{v}}_s$).

In the ballistic regime and in the transition between the two regimes, instead, the presence of $q_w$ has to be taken into account. In the experiments of Ref.~\cite{Bertman_Cry8(1968), Greywall_PRB23(1981)}  it occurs for temperatures below $1\operatorname{K}$ (diameter of the tubes are about  $10^{-3}\operatorname{m}$), but it depends strongly on the diameter of the channel, and for smaller channel it would occur for higher temperature. It is an interesting issue for future application in refrigeration of  nanosystems.

In rarefied phonon gas theory, or in rarefied particle gas, one finds that the slip heat flow along the walls is given as
	\be\label{q_w}
		q_w=-C\ell\left(\frac{\partial q}{\partial r}\right)_{wall}
	\ee
with $\ell$ being the mfp of the heat carriers (i.e., the phonons), and $C$ a non-negative numerical parameter (the values of which is smaller than the unit) which in a macroscopic formalism is an experimental coefficient, and which in kinetic theory of rarefied gases may be identified as $C=\left(2-f\right)/f$~\cite{Bertman_Cry8(1968), Greywall_PRB23(1981)}.

\subsection{Cylindrical channel}
Once the wall contribution $q_w$ is not negligible, but given by Eq.~\eqref{q_w}, following the same procedure adopted at the end of Sec.~\ref{section2}, namely, inserting Eq.~\eqref{eq:heat_flux3} into Eq.~\eqref{eq:heat_flux4}, one has
	\be\label{eq:heat_flux6}
		\dot Q l=\int_{T_1}^{T_2}\pi R^2\frac{R^2S^2 T}{8\eta}\left(1+4C\frac{\ell}{R}\right)\d T=\frac{\pi R^4  S^2 T_M}{8\eta}\left(1+4C\frac{\ell}{R}\right)	\Delta T,
	\ee
which yields the following effective thermal conductivity
	\be\label{eq:cond_1bis}
		K_{\text{eff}}=\frac{R^2 S^2 T_M}{8\eta}\left(1+4C\frac{\ell}{R}\right).
 	\ee

Whenever the ratio $\ell/R$ gets high values, Eq.~\eqref{eq:cond_1bis} becomes
	\be\label{eq:cond_1}
		K_{\text{eff}}=\frac{R^2S^2CT_M}{2\eta}\frac{\ell}{R}.
 	\ee

It is worth to note that the effective thermal conductivity~\eqref{eq:cond_1} reduces to the expression found in Ref.~\cite{Bertman_Cry8(1968)}, namely,
	\[
		k=\frac{1}{3} C_vvd\left(\frac{2-f}{f}\right)
	\]
by the following identifications
	\begin{subequations}\label{eq:subeq0}
		\begin{align}
    		&\eta=\frac{1}{5}\rho_nv\ell\label{eq:subeq1}\\
     		&S^2T=\frac{g\rho_n v^2C_v}{3} \label{eq:subeq3}
		\end{align}
	\end{subequations}
taken from Ref.~\cite{Greywall_PRB23(1981)}. In Ref.~\cite{Bertman_Cry8(1968)} it is also observed that nondimensional parameter $g$ is of the order of unity. Therefore, in  Eq.~\eqref{eq:subeq3} we are allowed to set $g=4/5$.

An expression analogous to Eq.~\eqref{eq:heat_flux6} was already used in the 1950's in Refs.~\cite{Bertman_Cry8(1968), Greywall_PRB23(1981), Maris_PRA7(1973), Benin_PRB18(1978)}. There, the authors took it directly on kinetic microscopical grounds, instead of using a non-local equation for heat transport complemented with a boundary condition. In Refs.~\cite{Bertman_Cry8(1968), Greywall_PRB23(1981)} the transition between the Landau regime to ballistic regime (between $0.7\operatorname{K}$ and $1\operatorname{K}$) is also complemented with the presence of rotons as heat carriers. For this reason, the entropy is the sum of the the entropy of rotons and phonons $S=S_{ph}+S_{rot}$; the mfp is due to the presence of the two kinds of carriers and for instance  one can use the one proposed by Khalatnikov\;\cite{Khalatnikov2000}.

In this case, from Eq.~\eqref{eq:heat_flux6} thermal conductivity  directly becomes
	\be\label{eq:cond_3}
		 K_{\text{eff}}=\frac{R^2S_{phon}^2T_M}{8\eta}\left(1+4C\frac{\bar\ell}{R}\right)\left(1+\frac{S_{rot}}{S_{phon}}\right)^2,
	\ee
wherein $\bar\ell$ is the mfp of the heat carriers (which in this case are both the phonons, and the rotons).

Note that thermal conductivity $\displaystyle k=\frac{R^2S_{phon}^2T}{8\eta}\left(1+4C\frac{\bar\ell}{R}\right)\left(1+\frac{S_{rot}}{S_{phon}}\right)^2$ in the integrand of Eq.~\eqref{eq:cond_3} coincides with Eq.~$(2)$ in Ref.~\cite{Bertman_Cry8(1968)} if Eqs.~\eqref{eq:subeq1} and~\eqref{eq:subeq3} hold together with the assumption that $\displaystyle C=\left(\frac{2}{3}\right)\frac{2-f}{f}$, and $\displaystyle\bar\ell=\frac{4}{3}\lambda$, where $\lambda$ is the mfp calculated by Khalatnikov~\cite{Khalatnikov2000}.

\subsection{Rectangular channel}
In a rectangular channel when the phonon mfp becomes of the same order of  the channel thickness $d$, the slip heat flux $q_w$ on the walls cannot be neglected in the heat-flux profile~\eqref{eq:soluz}. For it, the same definition of the previous sections  is used, namely
	\be\label{eq:heat_flux19}
		q_w=-C\ell \left(\frac{\d q\left(z\right)}{\d z}\right)_{|z=\pm d/2}=- \frac{C \ell d ST}{2\eta}\nabla p
	\ee
and the heat flow across the transversal section of sizes $a$ and $d$ in terms of $\nabla T$ is
	\be\label{eq:heat_flux20}
		\dot Q= -\frac{a S^2Td^3}{12 \eta}\left(1+\frac{6C\ell }{d}\right) \frac{\d T}{\d x},
	\ee

Following the same procedure as above, we find that the effective thermal conductivity is
\be\label{eq:conductivity1b}
		K_{\text{eff}}= \frac{d^2T_MS^2}{12\eta}\left(1+\frac{6C\ell }{d}\right).
	\ee

\section{The turbulent regime: transition to ballistic regime}\label{section4}
\setcounter{equation}{0}
In Section~\ref{section3} we have studied the transition from laminar regime to ballistic regime when the thickness  $d$ of the tube shrinks to the size of the phonon mfp. Since we are interested in the transition between the several possible regimes, and keping in mind that the mfp of vortex line is of the order of $L^{-1/2}$ then we will consider here the transition to the ballistic regime in  turbulent state, namely  $L^{-1/2}< \ell$ ($L$ being the vortex length density). 
 
In these cases, as in Ref.~\cite{Sciacca_2013}, we must consider an additional equation for the vortex line $L$:
	\be\label{equa_L}
		\frac{d L}{d t} =-\beta \kappa L^2+\left[\alpha_0 v_{ns}-\omega^\prime\beta\frac{\kappa}{d}\right] L^{3/2} 
	\ee
in such a way to have the following  equations in the steady state, generalizing Eqs.\eqref{eq:heat_flux2bis}:
    \begin{subequations}
        \begin{align}
            	\vspace{0.2cm}\displaystyle
        	&\nabla p-\frac{\eta}{S T}\nabla^2{\bf q}=0,\label{eq:heat_flux3a}\\
	        &S \nabla T-\nabla p+\frac{KSL}{\zeta} q=0,\label{eq:heat_flux3b}\\
		& -\beta \kappa L^2+\left[\alpha_0 v_{ns}-\omega^\prime\beta\frac{\kappa}{d}\right] L^{3/2} =0\label{eq:heat_flux3c}
        \end{align}
    \end{subequations}
wherein the coefficients $\alpha_0$ and $\omega^\prime$ are functions of $v_{ns}d/\kappa$, with $d$ being the smallest size of the tube (the diameter in the cylindrical channel and the distance between the plates in the rectangular channel), the coefficient $\beta$ takes into account of the destruction of vortices, and $\kappa=h/m$ is the quantum vorticity ($h$ means the Planck's constant, and $m$ the helium atomic mass).

Equation \eqref{eq:heat_flux3c} has the following steady-state solutions
    \begin{equation}\label{Lsolution}
        \begin{array}
            [c]{ll}
            L=0; & \displaystyle L^{1/2}=\frac{\alpha_0}{\beta \kappa}v_{ns}-\frac{\omega^\prime}{d}.
        \end{array}
    \end{equation}

The second solution is stable for $\displaystyle v_{ns}>V_{c1}=\frac{\beta \kappa\omega^\prime}{\alpha_0 d}$, and in Ref.~\cite{MartinTough} it is seen that it has two different regimes, namely a TI turbulence and TII turbulence flow. In the rectangular channel is not still clear the existence of the two states.

\subsection{Cylindrical channel}
In the same hypothesis of the previous section, namely, if the heat-flux depends only on the radius $r$ of the pipe, while the pressure and temperature depend instead on the position  $x$ along the axis of the channel, from Eq.~\eqref{eq:heat_flux3b} we obtain again that the heat flux profile in a generic section is given by Eq.~\eqref{eq:heat_flux3}
wherein the heat flux on the wall may be calculated by means of the constitutive equation~\eqref{q_w}, i.e.,
    \[
        \displaystyle q_w=q\left(r\right)_{r=R}=-C \ell\left(\frac{\pard q}{\pard r}\right)_{r=R}.
    \]

In particular, following the same steps as in the previous section, we firstly find
	\be\label{eq:heat_flux10}
		\bar q=-\frac{R^2 ST\nabla p}{8\eta}\left(1+\frac{4 C\ell}{R}\right)
	\ee
and then, replacing the value of $\nabla p$ in it with that arising from Eq.~\eqref{eq:heat_flux3b}, we finally obtain
	\be\label{eq:heat_flux11}
		\dot Q=-\pi R^2\left(\frac{A R^2 S}{1+A\frac{KS R^2\bar L}{\zeta}}\right)\nabla T
	\ee
wherein $\displaystyle A=\frac{ST}{8\eta}\left(1+\frac{4C \ell}{R}\right)$. In the previous expressions the terms with a bar refer to the mean value over the transversal section of the pipe.

Now the two different situations we are interested to take into account are: $L^{-1/2}< R \ll \ell$ and $L^{-1/2}\ll \ell \ll R$. The first situation includes the case in which $L^{-1/2}\ll R$ and the case $L^{-1/2}\approx  R$, which, according to results obtained by Martin and Tough~\cite{Toughbook1982, MartinTough}, they are  essentially related to the TI or TII turbulent regime. In both cases the mfp $\ell$ is longer than the intervortex space $L^{-1/2}$. This means that phonon-phonon collisions are very rare, but this does not give information on phonon-vortex collisions. This requires to revise the mfp $\ell$ and the coefficient $K$ of the mutual friction between vortices and heat carriers. Indeed, the actual value of the mfp $\ell$ will be given by $\ell=v/\tau$, with the relaxation time $\tau$ given by the Matthiessen's rule as
    \be
    \frac{1}{\tau}=\frac{1}{\tau_{phon-phon}}+\frac{1}{\tau_{phon-vort}}.
    \ee
wherein $\tau_{phon-phon}^{-1}$ is the frequency of phonon-phonon collisions, and $\tau_{phon-vort}^{-1}$ is the frequency of the phonon-vortex collisions. 

For large value of the phonon mfp, the coefficient $K$ (which is related to the friction between normal component and vortices), instead, becomes smaller and dependent of the probability about the interaction between heat carriers and vortices, namely $K\approx f\left(\frac{1}{\tau_{phon-vort}}\right)$.

\subsubsection*{\textbf{Case: $L^{-1/2}< R \ll \ell$}}
In this case the coefficient $\displaystyle A=\frac{C\ell ST}{2\eta R}$ and the heat flux $\dot Q$ in Eq.~\eqref{eq:heat_flux11} becomes
	\be\label{eq:heat_flux12}
		\dot Q=-\pi R^2\left(\frac{2\eta}{S^2 T R^2}+\frac{K \bar L}{\zeta}\right)^{-1}\nabla T
	\ee

Equation~\eqref{eq:heat_flux12} may be integrated along the tube in order to have
	\be\label{eq:heat_flux13}
		\dot Q l=\frac{\pi R^2 \zeta}{K \bar L}\Delta T-\frac{2\eta \zeta^2\pi R^2}{B K^2 \bar L^2}\ln\left(\frac{1+\frac{B KT_2 \bar L}{2\eta \zeta}}{1+\frac{B KT_1 \bar L}{2\eta \zeta}}\right)
	\ee
wherein $B=S^2 R C \ell$. If the temperature difference $\Delta T=T_2-T_1\ll T_1$, we are allowed to write
    \[
     \ln\left(\frac{1+\frac{B KT_2 \bar L}{2\eta \zeta}}{1+\frac{B KT_1 \bar L}{2\eta \zeta}}\right)\approx\left(1+\frac{B KT_1 \bar L}{2\eta \zeta}\right)^{-1}\Delta T
    \]
and hence
	\be\label{eq:heat_flux15}
		\dot Q l=\frac{\pi R^2 \zeta}{K \bar L}\Delta T-\frac{2\eta \zeta^2\pi R^2}{B K^2 \bar L^2}\left(1+\frac{B KT_1 \bar L}{2\eta \zeta}\right)^{-1}\Delta T.
	\ee

Since $K$ is related to the collisions of phonons against vortex lines, and since phonon collisions are very scarce (because $R\ll \ell$), it may presumed that $K$ will be negligible.

\subsubsection*{\textbf{Case: $L^{-1/2}\ll \ell \ll R$}}
In this case, the coefficient $A$ takes the value  $A=\displaystyle\frac{ST}{8\eta}$ and the heat flux $\dot Q$ in Eq.~\eqref{eq:heat_flux11} becomes
	\be\label{eq:heat_flux16}
		\dot Q=-\pi R^2\left(\frac{2\eta}{S^2 T R^2}+\frac{K \bar L}{\zeta}\right)^{-1}\nabla T
	\ee

Thus, Eqs.~\eqref{eq:heat_flux12} and~\eqref{eq:heat_flux16} coincide. The same integration along the tube that has been performed in Eqs.~\eqref{eq:heat_flux13} and~\eqref{eq:heat_flux15} could be done also here. However, in this case $K$ will be not negligible, in contrast to Eq.~\eqref{eq:heat_flux15}. Indeed, in this case the phonon-phonon collisions are more abundant than in the former case, and one could guess that also the phonon-vortex collisions will be more abundant, so that $K$ in this case will be higher than in the previous one.

\subsection{Rectangular channel}
In the transition to turbulence, as it was already pointed out in the section~\ref{section4}, Eqs.~\eqref{eq:heat_flux3a} and~\eqref{eq:heat_flux3b} have to be further complemented by Eq.~\eqref{eq:heat_flux3c} for the vortex line density $L$, having the steady state solutions~\eqref{Lsolution}.

The mean value of the heat flow $\bar q= \dot Q/(a d)$ can be obtained from the expression in Eqs.~\eqref{eq:soluz} and~\eqref{eq:heat_flux19}
	\be\label{eq:heat_flux21}
		\bar q= -\frac{STd^2}{12 \eta}\left(1+\frac{6C\ell }{d}\right) \nabla p,
	\ee
where we insert the expression of $\nabla p$ from  the averaged equation~\eqref{eq:heat_flux3b} over the transversal section. The results is
	\be\label{eq:heat_flux22}
		\dot Q= -\frac{Aad}{1+\frac{AKL}{\zeta}}\nabla T,
	\ee
wherein $\displaystyle A=\frac{S^2Td^2}{12 \eta}\left(1+\frac{6C\ell }{d}\right)$.

\section{Concluding remarks}\label{Concl}
\setcounter{equation}{0}
In this paper we have proposed a simple model for the transition of heat transfer in He II along channels (narrow cylindrical pipes and thin rectangular channel of high aspect ratio) in counterflow situation, from the Landau regime (with $\dot Q l/\Delta T\sim R^4$) to the ballistic regime (with $\dot Ql/\Delta T\sim R^3\ell$). Though the extreme situations, i.e., the pure Landau situation (with $\ell\ll R$) and the pure ballistic one (with $\ell\gg R$) are well-known, the transition between them is a topic of interest.

In contrast with the model proposed in Refs.~\cite{Bertman_Cry8(1968),Greywall_PRB23(1981)}, in the present paper we have complemented a macroscopic model based on the evolution equation for the local heat flux $\mathbf{q}$, with an evolution equation for $L$ and which may describe the transition from the Landau regime to the Gorter-Mellinck regime~\cite{Sciacca_2013}. We have also described the transition from turbulent regime to ballistic regime which  was not done before.  Thus, the model we have considered may describe the several regimes and the respective transitions amongst them.

This model for the transition has been based on a one-fluid model of He II with the heat  flux as independent variable. It has been shown that such model leads to many results which may be well-interpreted in the framework of the two-fluid model for He II. However, the ballistic situation is not properly analyzed in the two-fluid model, because here the viscous component is no longer so in the ballistic regime, where collisions with the walls are dominant over the bulk phonon-phonon collisions. In 
\cite{Cimmelli_PRB82(2010), Sellitto_JAP107(2010), AlvJouSel1} the similar problem of diffusive-to-ballistic transition was considered for solids by means of phonon-hydrodynamics. In this paper we have used a similar but more general formalism to dealt with heat transport in superfluid helium, where convective effect could be present, with their own viscous effects. 

The arguments of this paper could be useful for future applications in refrigeration of very small system by means of superfluid helium, where the ballistic contribution may be relevant. In Refs.~\cite{Bertman_Cry8(1968),Greywall_PRB23(1981)} it is stated that the transition to the ballistic regime occurs for temperatures lower than $1\operatorname{K}$ with diameter of the channels of the order of $10^{-3}\operatorname{m}$.  For narrower  channel (for instance to refrigerate nanosystems) the mfp of heat carriers (phonons) becomes of the same order of the system's size for  temperature higher than $1\operatorname{K}$.

\section*{Acknowledgements}
M.S. acknowledges the hospitality of the "Group of Fisica Estadistica of the Universit\`{a}t Aut\`{o}noma de Barcelona", the financial support of the Istituto Nazionale di Alta Matematica (GNFM--Gruppo Nazionale della Fisica Matematica) and  the support of the Universit\`{a}  di Palermo (under Grant Nos. Fondi 60\% 2012).

A.S.  acknowledges the financial support of the Italian GNFM---Gruppo Nazionale della Fisica Matematica --- under grant "Progetto Giovani 2012".

D.J. acknowledges the financial support from the Direcci\'{o}n General de Investigaci\'{o}n of the Spanish Ministry of Economy and Competitiveness under grant FIS2012-13370-C02-01 and of the Direcci\'{o} General de Recerca of the Generalitat of Catalonia, under grant 2009 SGR-00164 and to Consolider Programm Nanotherm.

\end{document}